\documentclass[a4paper]{jpconf}
\usepackage{graphicx}
\usepackage{cite}
\begin{document}

\hfill PITHA-07/17

\title{Recent developments in radiative $B$ decays}

\author{Tobias Huber}

\address{Institut f\"ur Theoretische Physik E, RWTH Aachen, D - 52056 Aachen, Germany}

\ead{thuber@physik.rwth-aachen.de}

\begin{abstract}
We report on recent theoretical progress in radiative $B$ decays. We focus on a calculation of logarithmically enhanced QED corrections to the branching ratio and forward-backward asymmetry in the inclusive rare decay $\bar B
\to X_s \ell^+ \ell^-$, and present the results of a detailed phenomenological analysis. We also report on the calculation of NNLO QCD corrections to the inclusive decay $\bar B \to X_s \gamma$. As far as exclusive modes are concerned
we consider transversity amplitudes and the impact of right-handed currents in the exclusive $\bar B \to K^\ast \ell^+ \ell^-$ decay. Finally, we state results for exclusive $B \to V \gamma$ decays, notably the time-dependent
CP-asymmetry in the exclusive $B \to K^\ast \gamma$ decay and its potential to serve as a so-called ``null test'' of the Standard Model, and the extraction of CKM and unitarity triangle parameters from $B \to (\rho,\omega) \gamma$ and
$B \to K^\ast \gamma$ decays.
\end{abstract}

\section{Introduction}
As flavor changing neutral current (FCNC) processes, radiative $b \to (s,d)$ transitions are very sensitive to physics beyond the Standard
Model (SM) since virtual effects due to heavy degrees of freedom are not overwhelmed by large tree-level contributions and hence their
impact is not necessarily suppressed with respect to the SM contributions. Therefore these transitions represent an ideal testing
ground for an indirect search for new physics (NP), and they are examined with unprecedented precision both on the theoretical and
experimental side. Remarkable results have been already achieved in this field, and more data still to come from the $B$ factories,
from LHC and from a possible SuperB factory is awaited with excitement.

\section{The inclusive decay $\bar B \to X_s \ell^+ \ell^-$}
\begin{figure}[t]
\begin{minipage}{18pc}

\vspace*{55pt}

\includegraphics[width=18pc]{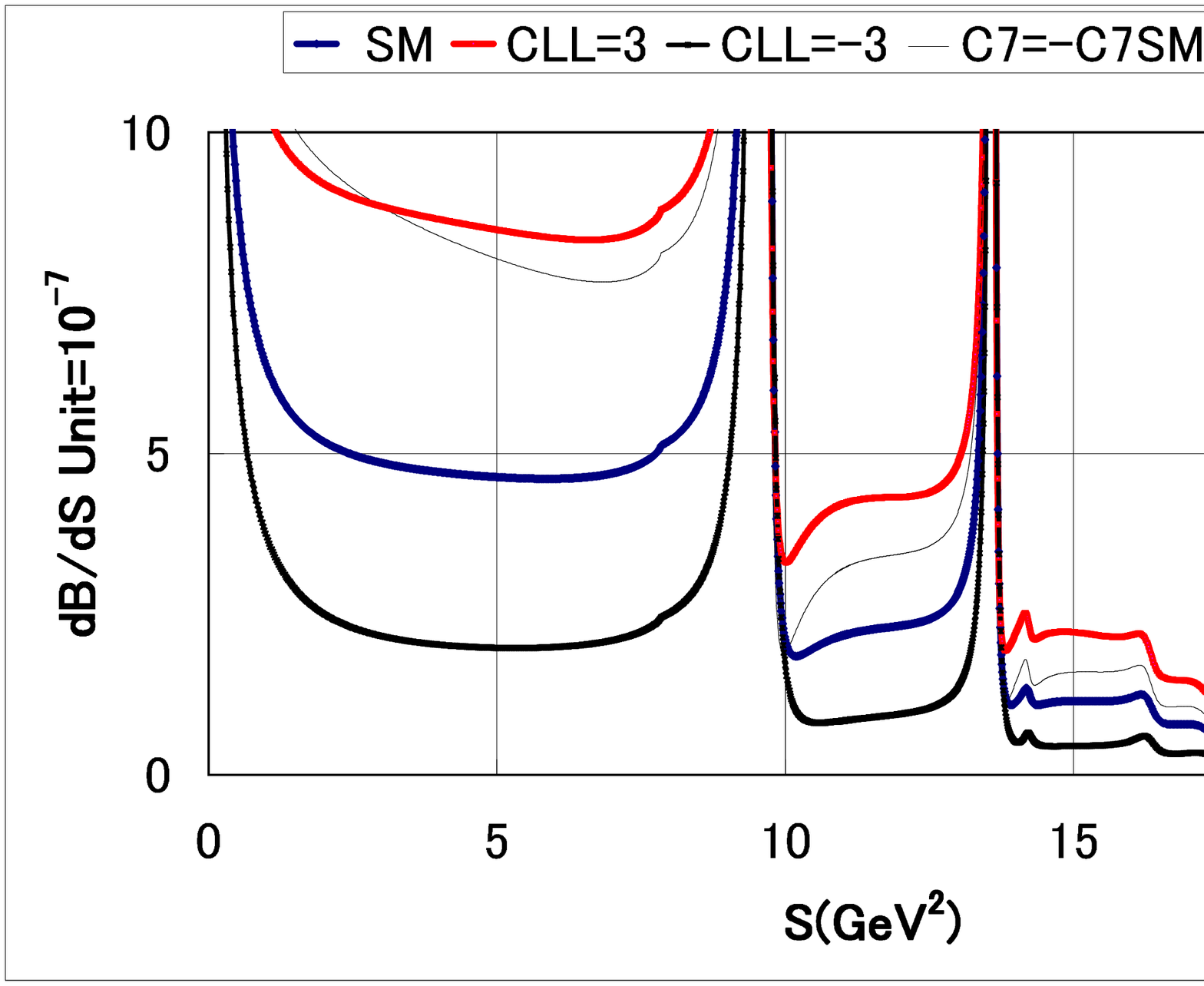}

\vspace*{-70pt}

\caption{\label{fig:br}Differential BR as a function of the lepton inv. mass. Second curve from bottom (blue): SM. Second curve from top (thin black): Sign of $C_7$ reversed w.r.t. SM. See~\cite{Akeroyd:2004mj} for more details.}
\end{minipage}\hspace{2pc}%
\begin{minipage}{18pc}
\includegraphics[width=18pc]{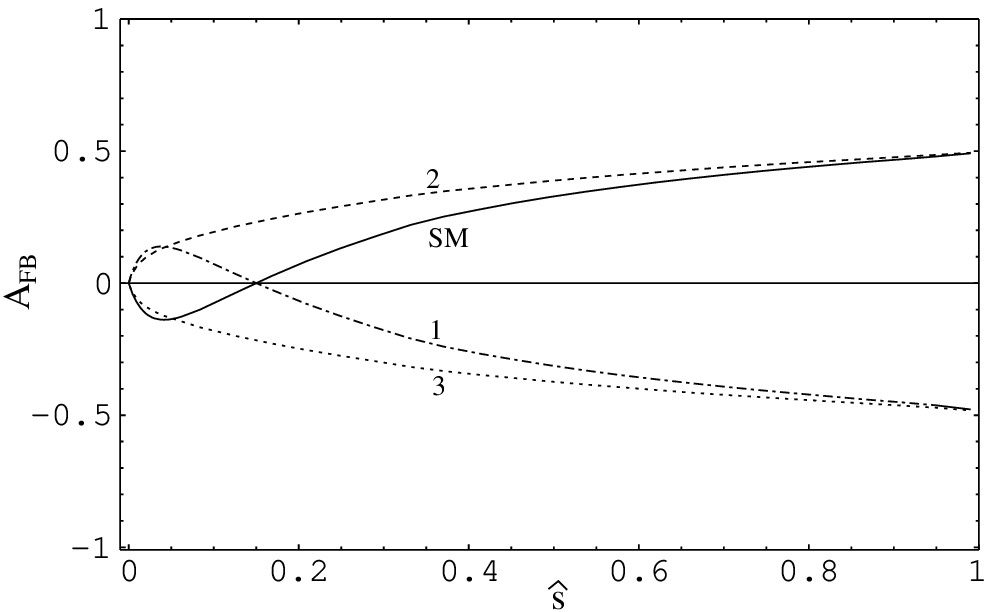}
\caption{\label{fig:fba}Forward backward asymmetry as a function of the lepton inv. mass. Curve 2: Reversed sign of $C_7$ w.r.t. SM.
Curves 1,3: Sign of $C_{10}$ reversed in addition to curves SM,2 respectively~\cite{Ali:2002jg}.}
%{also Eq.~(\ref{eq:fba})}
\end{minipage} 
\end{figure}

Alike many transitions in flavor physics, the dynamics of the decay $\bar B \to X_s \ell^+ \ell^-$ is most conveniently described by an
effective Hamiltonian with the top quark and the heavy gauge bosons being integrated out. Within this framework, occurring large
logarithms which stem from widely separated scales ${\cal O}(m_t,m_{W,Z})$ and ${\cal O}(m_b)$ can be efficiently resummed order by
order in perturbation theory. The corresponding effective operators and their associated Wilson coefficients can e.g.\ be found
in~\cite{Huber:2005ig}. The QCD corrections to the decay $\bar B \to X_s \ell^+ \ell^-$ have achieved NNLO 
accuracy~\cite{Asatryan:2001zw,Ghinculov:2002pe,Ghinculov:2003qd,Asatrian:2002va,Bobeth:1999mk,Bobeth:2003at}, and also
non-perturbative~\cite{Falk:1993dh,Buchalla:1997ky,Ali:1996bm,Kruger:1996cv,Chen:1997dj,Buchalla:1998mt,Bauer:1999kf} and higher order electroweak
corrections~\cite{Bobeth:2003at,Huber:2005ig} have been derived. The latter give rise to logarithmically enhanced corrections
proportional to $\alpha_{\rm em}\ln(m_b^2/m_{\ell}^2)$ which vanish upon integration over the entire phase space but are numerically relevant 
if one is restricted to certain regions of the invariant mass $q^2$ of the leptons~\cite{Huber:2005ig}. The two most important quantities in
$\bar B \to X_s \ell^+ \ell^-$ are the differential branching ratio (BR) and forward backward asymmetry (FBA). In particular, the
value $q_0^2$ for which the differential FBA vanishes is one of the most precise predictions in flavor physics with a theoretical
uncertainty of order 5\%. A thorough phenomenological analysis which includes all known corrections yields for the BR integrated over
the low-$q^2$ ($q^2 \in [1,6]$~GeV$^2$) region~\cite{Huber:2005ig}
\begin{equation}
{\cal B}(\bar B \to X_s \mu \mu)=(1.59 \pm 0.11)\cdot 10^{-6} \, ,
\end{equation}
 where the indicated uncertainty includes only the parametric and
perturbative ones. No additional uncertainty for the unknown subleading
non-perturbative corrections has been included. In particular, 
the uncalculated ${\cal O}(\alpha_s(\mu_b) \, \Lambda_{\rm QCD}/m_{c,b})$ 
non-perturbative corrections imply an additional uncertainty of around 
$ 5\%$ in the above formula. 
The current experimental world average of this quantity is 
$(1.60 \pm 0.51)\cdot 10^{-6}$~\cite{Iwasaki:2005sy,Aubert:2004it}. For the BR integrated over the high-$q^2$ ($q^2>14.4$~GeV$^2$)
region one obtains~\cite{Huber:2008hl}
\begin{equation}
{\cal B}(\bar B \to X_s \mu \mu)=(2.40^{+0.69}_{-0.62})\cdot 10^{-7} \, .
\end{equation}
The corresponding experimental
values are $(4.18 \pm 1.17_{stat} \, ^{+0.61}_{-0.68} \, \! _{sys.})\cdot 10^{-7}$~\cite{Iwasaki:2005sy} and \\$(5 \pm 2.5_{stat.} \,
^{+0.8}_{-0.7} \, \!_{sys.})\cdot 10^{-7}$~\cite{Aubert:2004it}. As far as the FBA is concerned, we find for the position $q_0^2$ for
which the FBA vanishes~\cite{Huber:2008hl}
\begin{equation}
q^2_{0,\mu\mu}=(3.50 \pm 0.12) {\rm GeV}^2 \, .
\end{equation}
By the end of the $B$ factories the fully
differential shape of the BR and FBA will not be accessible, contrary to their integrals over one or two bins in the low-$q^2$ region. However,
these quantities will already allow to discriminate between different NP scenarios (see Figures~\ref{fig:br} and~\ref{fig:fba} as well as Ref.~\cite{Gambino:2004mv}), and their SM
predictions are given in Ref.~\cite{Huber:2008hl}. Quite recently two additional quantities related to $\bar
B \to X_s \ell^+ \ell^-$ have been proposed. One is related to the structure of the double differential
decay width and represents a third linearly independent quantity in addition to the BR and FBA~\cite{Lee:2006gs}. The other one is the differential decay
width integrated over the high-$q^2$ region and normalized to the semileptonic $b \to u \ell \nu$ rate {\textit{with the same
cut}}~\cite{Ligeti:2007sn} in order to significantly reduce the error due to parameters in the non-perturbative $1/m_b$
corrections~\cite{Ligeti:2007sn,Huber:2008hl}.
Moreover, an experimental determination of this quantity might become feasible by the end of the $B$ factories.
\section{The inclusive decay $\bar B \to X_s \gamma$}
The inclusive rare decay $\bar B \to X_s \gamma$ is the most prominent among the radiative $b \to s$ transitions. Its branching ratio has been measured at several accelerator
facilities~\cite{Chen:2001fja,Abe:2001hk,Koppenburg:2004fz,Aubert:2005cua,Aubert:2006gg}, yielding a world average of~\cite{Barberio:2006bi} ${\cal B}(\bar B \to X_s \gamma)^{\rm{exp.}} = (3.55 \pm
0.24^{+0.09}_{-0.10}\pm 0.03) \times 10^{-4}$ for a photon energy cut of $E_{\rm cut}>1.6$~GeV in the restframe of the $\bar B$. The errors are combined statistical and systematic, systematic due to extrapolation in $E_{\rm cut}$, and
due to the $b \to d \gamma$ fraction. By the end of the $B$ factories these errors are expected to decrease to around 5\% due to larger statistics and possible lower cuts on $E_{\gamma}$. This also calls for precise predictions on the
theoretical side. Alike in $\bar B \to X_s \ell^+ \ell^-$ this decay is described in the framework of the effective hamiltonian. The different steps of calculating matching conditions for the Wilson coefficients at the electroweak
scale~\cite{Bobeth:1999mk,Misiak:2004ew}, determining the anomalous dimensions matrix which governs the running of the Wilson
coefficients~\cite{Gorbahn:2004my,Gorbahn:2005sa,Czakon:2006ss}, and finally the extraction of on-shell matrix
elements~\cite{Blokland:2005uk,Melnikov:2005bx,Asatrian:2006ph,Asatrian:2006sm,Asatrian:2006rq,Bieri:2003ue,Misiak:2006ab,Boughezal:2007ny}, have now
achieved the NNLO level and involve multi (two to four)-loop
calculations. The resulting SM prediction, which includes also electroweak corrections~\cite{Czarnecki:1998tn,Kagan:1998ym,Baranowski:1999tq,Gambino:2000fz,Gambino:2001au} as well as non-perturbative $\Lambda^2/m^2_{b,c}$
corrections~\cite{Chay:1990da,Khodjamirian:1997tg,Voloshin:1996gw,Bauer:1997fe,Ligeti:1997tc,Grant:1997ec,Buchalla:1997ky,Neubert:2004dd}, is
$
\displaystyle{\cal B}(\bar B \to X_s \gamma)_{\rm{SM}}^{E_\gamma>1.6\rm{ GeV}} = (3.15 \pm 0.23) \times 10^{-4}
$~\cite{Misiak:2006zs}.
The unknown ${\cal O}(\alpha_s
\Lambda/m_b)$ corrections are estimated to be of order 5\%~\cite{Misiak:2006zs,Lee:2006wn}. The other uncertainties which contribute to the total error are parametric uncertainties (3\%), scale uncertainties (3\%),
and an uncertainty (3\%) due to an interpolation in $m_c$ in the computation of the three-loop matrix elements of $P_{1,2}$. In going from NLO to NNLO accuracy, the scale dependence gets tremendously reduced, which is in particular
true for the charm scale $\mu_c$. This scale first enters at NLO and hence one needs NNLO precision in order to tame its dependence~\cite{Misiak:2006zs}. %In another recent analysis, the authors of Ref.~\cite{Becher:2006pu} perform a
%multiscale operator product expansion (MSOPE) with three short-distance scales ($m_b$,$\, \Delta=m_b-2E_{\gamma}$,$\, \sqrt{m_b\Delta}$) to connect shape function and local OPE region. Using soft-collinear effective theory (SCET),
%their analysis finds a value of which is around 5\% lower compared to the one given above. The achieved theoretical and experimental precision now allows one to constrain parameter spaces of new physics (NP)
%models. Examples of recent analyses are the lower bound on the mass of the $H^\pm$ in type II 2HDM which is found to be $M_{H^\pm} > 295$~GeV at 95\% CL independent of $\tan\beta$~\cite{Misiak:2006zs}. Furthermore, an analysis in
%minimal universal extra dimensions (mUED) finds a lower bound on the inverse compactification scale $1/R > 600$~GeV at 95\% CL.
For other recent work on this decay mode see Refs.~\cite{Hurth:2007xa,Haisch:2007ic} and references therein.
\section{Transversity amplitudes in $B \to K^\ast (K \pi) \ell^+\ell^-$}
The matrix element of the $B \to K^\ast$ transition can be parameterized in terms of seven a priori
independent form factors. However, in the limit of a heavy quark and a large $E_{K^\ast}$ the seven $B \to K^\ast$ form factors reduce to two universal ones~\cite{Charles:1998dr,Beneke:2000wa}.
Those form factors in turn cancel out in specific transverse asymmetries, which then depend on short-distance information only~\cite{Kruger:2005ep,Lunghi:2006hc}, and SM prediction on these 
transverse asymmetries can be found in Refs.~\cite{Kruger:2005ep,Lunghi:2006hc}. 
%\begin{equation}
%\displaystyle A_{T}^{(1)}(s) = \frac{-2 {\rm Re}(A_{\parallel}\, A_{\perp}^\ast)}{\left|A_{\perp}\right|^2+\big|A_{\parallel}\big|^2} \, ,
%\qquad A_{T}^{(2)}(s) =
%\frac{\left|A_{\perp}\right|^2-\big|A_{\parallel}\big|^2}{\left|A_{\perp}\right|^2+\big|A_{\parallel}\big|^2} \, .
%\end{equation}
%Including next-to-leading order corrections and integrating over the low di-muon mass region $2 \, m_{\mu} \le M_{\mu\mu} \le 2.5$ GeV (but
%leaving out $\Lambda_{\rm QCD}/m_b$ corrections), one finds
%$
%\displaystyle A_{T}^{(1)} = 0.9986 \pm 0.0002$, $\quad A_{T}^{(2)} = - 0.0043 \pm 0.003 \,
%$~\cite{Kruger:2005ep}.
However, there are NP scenarios where there can be huge deviations from the SM values of these asymmetries. In Ref.~\cite{Lunghi:2006hc} one can find examples for
the case of the MSSM with R-parity and non-MFV in down-squarks soft-breaking terms. Transverse asymmetries therefore provide a
theoretically clean way to analyse the chiral structure of the $b \to s$ current. For another very recent analysis on angular distributions
in $\bar B \to K \ell^+\ell^-$ see Ref.~\cite{Bobeth:2007dw}.
\section{Time-dependent CP-asymmetry in $B \to K^\ast \gamma$}
The time-dependent CP-asymmetry (TDCPA) in $B \to K^\ast \gamma$~\cite{Atwood:1997zr},
\begin{equation}\label{eq:TDCPA}
\displaystyle A_{\rm CP} (t)= \frac{\Gamma(\bar B^0(t)\to \bar K^{\ast 0} \gamma)-\Gamma(B^0(t)\to K^{\ast 0} \gamma)}{\Gamma(\bar B^0(t)\to \bar
K^{\ast 0} \gamma)+\Gamma(B^0(t)\to K^{\ast 0} \gamma)}= S \, \sin(\Delta m_B \, t) - C \, \cos(\Delta m_B \, t) \, ,
\end{equation}
is believed to be small in the SM. Due to the operator
$
\displaystyle Q_7 = \bar s \, \sigma^{\mu\nu} \, F_{\mu\nu} \, (m_b \, P_R + m_s \, P_L) \, b \, ,
$
which follows from the structure of the weak interaction, the emitted photon is predominantly left-handed in $b$ and right-handed in $\bar b$ decays. Hence the TDCPA is suppressed by a factor of ${\cal O}(m_{s,d}/m_b)$. On the other
hand, the TDCPA can be enhanced by terms of ${\cal O}(m_{\rm heavy}/m_b)$ due to a helicity flip on heavy internal
lines in NP models. %such as left-right symmetric models~\cite{Mohapatra:1974hk,Mohapatra:1974gc,Senjanovic:1975rk,Senjanovic:1978ev,Atwood:1997zr,Grinstein:2004uu,Grinstein:2005nu},
%supersymmetry~\cite{Frank:2002nj,Chun:2000hj,Everett:2001yy,Goto:2003iu,Chua:2003xq,Hou:2006mz}, warped extra dimensions~\cite{Agashe:2004ay,Agashe:2004cp}, and anomalous right-handed top quark couplings~\cite{Lee:2003ci}.
This quantity is therefore considered a prime candidate for a so-called ``null-test'' of the SM~\cite{Gershon:2006mt}. However, there is a possible enhancement~\cite{Grinstein:2004uu,Grinstein:2005nu} also in the SM due to gluon
emission from a quark loop generated by operators like $Q_2 = (\bar c \, \gamma^{\mu} \, P_L \, b) (\bar s \, \gamma_{\mu} \, P_L \,
c)$~\cite{Ali:1990tj,Pott:1995if}. 
%The resulting suppression factor is ${\cal O}(\alpha_s)$ in inclusive and ${\cal O}(\Lambda_{\rm QCD}/m_b)$ in exclusive decays, both for $b \to s \gamma$ and $b \to d \gamma$~\cite{Grinstein:2004uu,Grinstein:2005nu}.
If these contributions turn out to be small, one can interpret a possible large value of the TDCPA as
a signal for NP~\cite{Ball:2006eu}. A value for $S(B
\to K^\ast \gamma)$ in Eq.~(\ref{eq:TDCPA}) has been derived at several places in the literature.
The analyses in Refs.~\cite{Ball:2006cva,Ball:2006eu}, which combine QCD-factorisation with QCD sum rules on the light-cone to estimate long-distance photon emission and soft-gluon
emission from quark loops yield
$ \displaystyle
S = - 0.022 \pm 0.015^{+0}_{-0.1} \quad {\rm and} \quad S\left|_{\rm soft \; gluons} \right. = 0.005 \pm 0.01 \, ,
$
whereas a conservative dimensional estimate (from a SCET based analysis) gives
$
\big|S_{\left|_{\rm soft \; gluons} \right.}\big| \approx 0.06 \, 
$~\cite{Grinstein:2004uu,Grinstein:2005nu}.
There are, however, arguments that for the $B \to K^\ast \gamma$ channel the number extracted in Refs.~\cite{Grinstein:2004uu,Grinstein:2005nu} can be smaller, see Ref.~\cite{Hurth:2007xa} and references therein for a recent
discussion. The calculation in pQCD yields $\displaystyle S_{\rm pQCD} = -0.035\pm 0.017$~\cite{Matsumori:2005ax}, where the effect is mainly from hard gluons, and soft ones are treated in a model dependent way. The experimental world
average reads~\cite{Aubert:2005bu,Ushiroda:2006fi,Barberio:2006bi} $ S_{\rm HFAG}= -0.28\pm 0.26$.
While LHC will have better performance in decays like $B_s \to \phi \gamma$, a SuperB factory can measure $S(B \to K^\ast \gamma)$ with an uncertainty as low as $0.04$ at $50$ab$^{-1}$~\cite{Akeroyd:2004mj}.
%Consider TPCPA instead of merely BR's since the helicity amplitudes add incoherently in BR's but coherently in TDCPA. Hence a large NP effect could escape detection if only BR's were considered.\\
%TDCPA in $\bar B^0 \to V \gamma$ is small in the SM, irrespective of hadronic interactions. NP effects relief this suppression. \\
\section{Extraction of CKM and UT parameters from $B \to (\rho,\omega) \gamma$ and $B \to K^\ast \gamma$ decays}
We finally would like to report on an analysis which was performed in Ref.~\cite{Ball:2006eu}. One considers {\textit{ratios}} of branching ratios (BR) of exclusive radiative $B$ decays since the ratios
of the occurring form factors are much better known than the individual form
factors themselves. %The absolute scale for the BRs -- i.e.\ the BR to which one normalizes -- is then set by the one with the smallest experimental uncertainty, which is in this case the CP and isospin averaged BR $\bar {\cal B}(B \to
%K^\ast \gamma)$.
The following two observables are particularly interesting,
$
 \displaystyle \quad R_{\rho/\omega} \equiv \bar{\cal B}(B \to (\rho,\omega) \gamma)/\bar{\cal B}(B \to
K^\ast \gamma),$$\quad R_{\rho} \equiv \bar{\cal B}(B \to \rho \gamma)/\bar{\cal B}(B \to K^\ast \gamma),
$
where the BRs are CP and isospin averaged. The knowledge of these two quantities --- and a few other parameters --- allows one to extract $|V_{td}/V_{ts}|$ as well as the UT angle $\gamma$, where the extraction of the latter
involves a twofold degeneracy $2\pi \leftrightarrow 2\pi-\gamma$. The extraction of $\gamma$ from tree-level CP asymmetries in $B \to D^{(\ast)}
K^{(\ast)}$~\cite{Poluektov:2006ia} on the other hand carries a twofold degeneracy $\pi \leftrightarrow \pi+\gamma$. Combining these two different degeneracies hence allows one to {\textit{unambiguously}} determine the UT angle
$\gamma$. With the most recent results from the $B$-factories~\cite{Aubert:2006ag,Aubert:2006pu,Abe:2005rj,Barberio:2006bi} for the
above ratios, the authors of Ref.~\cite{Ball:2006eu} find, under the assumption of a unitary CKM matrix, that the solution $\gamma \, <
\, 180^\circ$ is clearly favored,
\begin{eqnarray}
\displaystyle {\rm BaBar:} \quad \left|V_{td}/V_{ts}\right| = 0.199^{+0.022}_{-0.025}({\rm exp}) \pm 0.014({\rm th}) &\leftrightarrow& \quad \gamma =
(61.0^{+13.5}_{-16.0}({\rm exp})^{+8.9}_{-9.3}({\rm th}))^\circ \nonumber \\
\displaystyle {\rm Belle:} \quad \left|V_{td}/V_{ts}\right| = 0.207^{+0.028}_{-0.033}({\rm exp})^{+0.014}_{-0.015}({\rm th}) \quad &\leftrightarrow& \quad \gamma
= (65.7^{+17.3}_{-20.7}({\rm exp})^{+8.9}_{-9.2}({\rm th}))^\circ.
\end{eqnarray}
%%%%%%%%%%%%%%%%%%%%%%%%%%%%%%%%%%%%%%%%%%%%%%%%%%%%%%%%%%%%%%%%%%%%%%%%%%%%%%%%%%%%%%%%%%%%%%%%%%%%%%%%%%%%%%%%%%%%%%%%%
%%%%%%%%%%%%%%%%%%%%%%%%%%%%%%%%%%%%%%%%%%%%%%%%%%%%%%%%%%%%%%%%%%%%%%%%%%%%%%%%%%%%%%%%%%%%%%%%%%%%%%%%%%%%%%%%%%%%%%%%%
\ack

I would like to thank the organizers of EPS 2007 and the convenors of the flavor session for the excellent organization of the
conference and for creating an inspiring atmosphere. I am greatful to Tobias Hurth, Enrico Lunghi, and Miko{\l}aj Misiak for a careful
reading of the manuscript and for valuable comments. This work was supported by Deutsche Forschungsgemeinschaft SFB/TR 9
``Computergest\"utzte Theoretische Teilchenphysik''.

\section*{References}
%\bibliographystyle{iopart-num}
%\bibliography{iopart-num}
\providecommand{\newblock}{}

\end{document}